\documentclass[aps,twocolumn,superscriptaddress]{revtex4-1}
\usepackage{amsmath,amsthm,verbatim} %amsrefs
\usepackage{amssymb}
\usepackage{graphicx}
\usepackage{bbm,bm}
\usepackage[caption=false]{subfig}
\usepackage[rgb]{xcolor}
\usepackage{mathrsfs}
\usepackage{stackrel} % adds [below] option to \stackrel
\usepackage[all]{xy}
\usepackage{float} % provides the 'H' option of the figure environment

\usepackage{natbib}

\usepackage{enumerate,amsthm,amsmath,amssymb,color,graphicx,bbm,float,framed}

\newcommand{\beq}{\begin{equation}}
\newcommand{\eeq}{\end{equation}}

\newcommand{\tr}{\mathrm{tr}}

\definecolor{myurlcolor}{rgb}{0,0,0.4}
\definecolor{mycitecolor}{rgb}{0,0.5,0}
\definecolor{myrefcolor}{rgb}{0.5,0,0}
\usepackage{hyperref}
\hypersetup{colorlinks,
linkcolor=myrefcolor,
citecolor=mycitecolor,
urlcolor=myurlcolor}

\makeatletter
\newtheorem*{rep@theorem}{\rep@title}
\newcommand{\newreptheorem}[2]{%
\newenvironment{rep#1}[1]{%
 \def\rep@title{#2 \ref{##1}}%
 \begin{rep@theorem}}%
 {\end{rep@theorem}}}
\makeatother

\newreptheorem{theorem}{Theorem}

\newreptheorem{lemma}{Lemma}

\usepackage{mathtools}
\usepackage[normalem]{ulem}

\begin{document}

\title{Asymptotic security of continuous-variable quantum key distribution\\ with a discrete modulation}

\author{Shouvik Ghorai}
\affiliation{LIP6, CNRS, Sorbonne Universit\'{e}, Paris, France}
\author{Philippe Grangier}
\affiliation{Laboratoire Charles Fabry, IOGS, CNRS, Universit\'e Paris Saclay, F91127 Palaiseau, France}
\author{Eleni Diamanti}
\affiliation{LIP6, CNRS, Sorbonne Universit\'{e}, Paris, France}
\author{Anthony Leverrier}
\affiliation{Inria Paris, France}

\date{\today}

\begin{abstract}
We establish a lower bound on the asymptotic secret key rate of continuous-variable quantum key distribution with a discrete modulation of coherent states. The bound is valid against collective attacks and is obtained by formulating the problem as a semidefinite program. We illustrate our general approach with the quadrature phase-shift keying (QPSK) modulation scheme and show that distances over 100 km are achievable for realistic values of noise. We also discuss the application to more complex quadrature amplitude modulations (QAM) schemes. 
This result opens the way to establishing the full security of continuous-variable protocols with a discrete modulation, and thereby to the large-scale deployment of these protocols for quantum key distribution.
\end{abstract}

\maketitle

%%%%%%%%%%%

Quantum key distribution (QKD) is the task of establishing a secret key between two distant parties, Alice and Bob, who can access an untrusted quantum channel and an authenticated classical channel \cite{SBC08}. Remarkably, very simple protocols based on the exchange of quantum states exist and have been proven secure against any eavesdropper only limited by the laws of quantum mechanics. The first QKD protocol, BB84, was invented by Bennett and Brassard and simply requires Alice to send qubit states from the set $\{|0\rangle, |1\rangle, |+\rangle = \frac{1}{\sqrt{2}} (|0\rangle+|1\rangle), |-\rangle = \frac{1}{\sqrt{2}} (|0\rangle-|1\rangle)\}$ through the quantum channel, and Bob to perform a measurement in one of the two bases $\{|0\rangle, |1\rangle\}$ or $\{|+\rangle, |-\rangle\}$. This provides them with some correlated data which can then be distilled into a secret key, provided that the correlations are large enough \cite{BB84}. 

The main drawback of BB84-like protocols based on the exchange of qubit states lies in the detection part which necessitates single-photon detectors. An interesting solution to avoid this costly and specific equipment is to replace it by coherent detection, which is the current industry standard in coherent optical telecommunication \cite{GJR17,JEM18}. 
This is the idea behind continuous-variable (CV) QKD \cite{ral99, GG02, GVW03}. 
In CVQKD protocols, information is encoded on the quadratures of the quantized electromagnetic field: Alice prepares coherent states, \textit{i.e.}, displaced vacuum states, while Bob performs homodyne or heterodyne (also called double-homodyne) detection to establish some correlations with Alice \cite{DL15}. These correlations can then be turned into a secret key by a classical postprocessing procedure similar to that of BB84.

Continuous variables enjoy a number of advantages for QKD: the hardware implementation is simpler since it corresponds to techniques already deployed in classical telecommunication and the secret key rate (\textit{i.e.}, the ratio between the final key size and the number of states exchanged on the quantum channel) is higher than for qubit-based protocols \cite{PLO17}. In fact, the main difficulty arising with CVQKD concerns security proofs: because the description of the protocol explicitly involves an infinite dimensional Fock space, many of the proof techniques developed for qubit-based protocols become unavailable, and new approaches are needed. 
 
The Graal in the context of security proofs is to establish a composable security proof in the finite-size regime, valid against general attacks. For BB84, it took about 20 years to reach that level, most notably with the work of Renner \cite{ren08} and better analyses continue to improve the key rates \cite{TR11, TLG12, DFR16, TL17, ADF18}.
The situation is less advanced for CVQKD since only a few CV protocols are currently known to enjoy such security: protocols based on the exchange of coherent states and heterodyne detection \cite{lev15, lev17, GDL19}, and protocols with squeezed states and homodyne detection \cite{FFB12, fur14}, but crucially only protocols where the states are modulated according to a \emph{Gaussian} distribution.
This state-of-affairs is not quite satisfactory because a Gaussian modulation can never be perfectly achieved in practice, and real protocols necessarily approximate such a Gaussian by some finite constellation of finite energy \cite{JKD12, KGW19}. Beyond this theoretical argument, a discrete modulation would present important advantages both on the hardware side since it would simplify the state preparation procedure \cite{ral99, rei00, HYA03, LKL04} and on the software side since the crucial step of error correction is dramatically simplified with a small constellation of states \cite{LG09}. More generally, if quantum key distribution is to be deployed at large scale, it is crucial that it conforms as much as possible to telecom standards, which currently involve discrete modulations of coherent states and coherent detection.

For these reasons, an outstanding and pressing open question of the field is to establish the security of CVQKD with a discrete modulation. Current security proofs restrict the possible attacks performed by the eavesdropper to emulate a linear quantum channel between Alice and Bob \cite{LG09} (see also \cite{HL07} and \cite{SL10}). We also note that Ref.~\cite{ZHR09} analyzed the security of a 2-state protocol and \cite{BW18} the security of a 3-state protocol, but the corresponding bounds are very pessimistic in term of resistance to loss, and the proof techniques in these papers are unlikely to easily generalize to more useful modulation schemes.
An alternative approach to simplify the error correction procedure is to rely on post-selection  \cite{SRL02,LKL04,LSS05}, but security proofs for such protocols are currently restricted to Gaussian attacks, which are not believed to be optimal \cite{HL07,SAA07}. Gaussian post-selection has also been investigated in the literature mainly because security proofs are easier to obtain \cite{FC12, WRS13}, but the performance of these variants is still not well understood.

In this paper, we present a major step towards the full security of CVQKD with a discrete modulation, by introducing a new proof technique that establishes a lower bound valid against arbitrary collective attacks, in the asymptotic limit of infinitely long keys. For concreteness, we first illustrate it for the quadrature phase-shift keying (QPSK) protocol and then discuss its extension to larger quadrature amplitude modulations (QAM). 
This is significant since the secret key rate against collective attacks, where the quantum channel is assumed to be identical for all uses, usually coincides with the secret key rate valid against arbitrary attacks in the asymptotic limit \cite{ren07, RC09, lev17}. Obtaining a composable security proof as well as computing the secret key rate in the finite-size regime would require to fully address the parameter estimation procedure, which is left for future work.

The outline of the paper is as follows. In Section \ref{sec:prot}, we recall the description of the QPSK protocol of \cite{LG09}. In Section \ref{sec:chal}, we discuss the specific challenges raised by the security analysis of CVQKD protocols with a discrete modulation. We present our security proof for the QPSK protocol and some numerical results in Section \ref{sec:bound}, and then explain how to extend the approach to more general QAM in Section \ref{sec:larger}. We finally discuss some limitations and outline future work in Section \ref{sec:disc}.

%%%%%%%%%%%%%%%%%%

\section{The QPSK protocol}
\label{sec:prot}

The QPSK constellation that we consider consists of four coherent states $\{|\alpha_k\rangle\}_{k=0\cdots 3}$ with $|\alpha_k\rangle := |i^k \alpha\rangle = e^{-\alpha^2/2} \sum_{n\geq 0} e^{i kn\frac{\pi}{2}}\frac{\alpha^n}{\sqrt{n!}} |n\rangle$, where $\alpha >0$ is a parameter to be optimized later. 
\begin{figure}[!h]
\centering
  \includegraphics[width=0.8\linewidth]{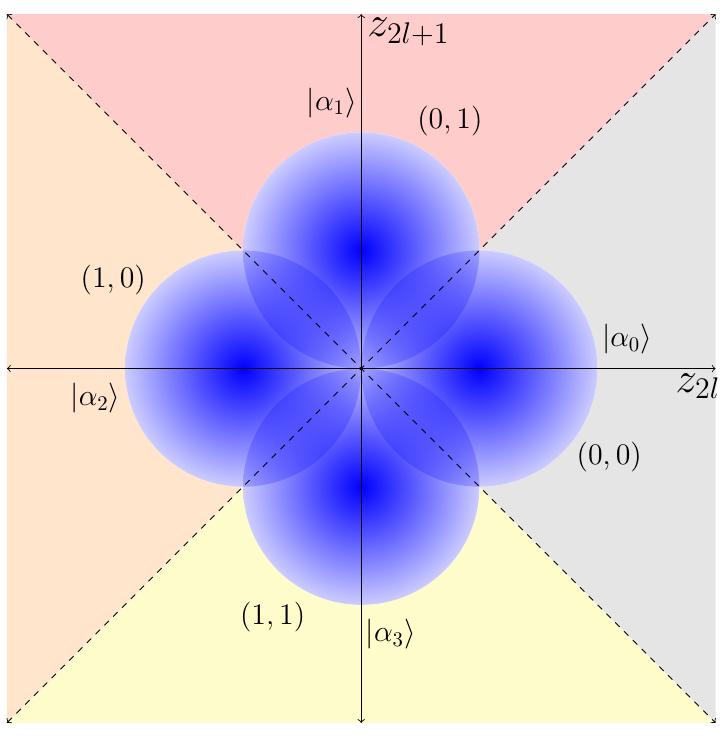}
\caption{Description of the QPSK protocol with a constellation of 4 coherent states and the partition of phase-space in four quadrants.}  
  \label{fig:prot}
\end{figure}
The Prepare-and-Measure (PM) version protocol is as follows. Alice picks a random bit string $\mathbf{x} = (x_0, \ldots, x_{2L-1})$ of length $2L$ (for some large $L$) and successive pairs of bits are encoded as coherent states of the form $|\alpha_{k_\ell}\rangle$ with $k_\ell = 2x_{2\ell} + x_{2\ell+1}$, as depicted in Fig.~\ref{fig:prot}. She sends these coherent states through the channel and Bob measures each output mode with heterodyne detection to obtain a $2L$-string $\mathbf{z} = (z_0, \ldots, z_{2L-1}) \in \mathbbm{R}^{2L}$. This string is then converted into a raw key of $2L$ bits $\mathbf{y} = (y_0, \ldots, y_{2L-1})$ given by
\begin{align*}
(y_{2\ell}, y_{2\ell+1}) = \left\{
\begin{array}{l}
(0,0) \quad \text{if} \quad z_{2\ell+1} <z_{2\ell},  \quad z_{2\ell+1}\geq -z_{2\ell},\\
(0,1) \quad \text{if} \quad z_{2\ell+1} \geq  z_{2\ell}, \quad z_{2\ell+1}> -z_{2\ell},\\
(1,0) \quad \text{if} \quad z_{2\ell+1} > z_{2\ell}, \quad z_{2\ell+1}\leq-z_{2\ell},\\
(1,1) \quad \text{if} \quad z_{2\ell+1} \leq z_{2\ell}, \quad z_{2\ell+1}< -z_{2\ell}.
\end{array}
\right.
\end{align*}
Bob further reveals the values of $|z_{2 \ell} \pm z_{2\ell+1}|$ publicly. This information allows Alice and Bob to turn the information reconciliation problem into a well-studied channel coding problem for the binary-input additive white-noise Gaussian channel (see Section 5.2 of \cite{Lev09} for further details about this procedure). The remaining steps of the protocol are standard, namely parameter estimation (discussed below), information reconciliation (Bob sends additional information on the classical channel to help Alice guess the string $\mathbf{y}$) and privacy amplification (so that Eve has no information about the final key).

The goal of parameter estimation is to decide whether the raw key can be turned into a secret key via classical postprocessing. More precisely, the idea is to check that the correlations between Alice and Bob's strings $\mathbf{x}$ and $\mathbf{z}$ are strong enough to guarantee that Eve only has limited knowledge about the raw key.
For BB84-like protocols, parameter estimation consists in evaluating the quantum bit error rate between Alice and Bob's data. For CVQKD, the parameter of interest is the \textit{covariance matrix}. In particular, for the 4-state protocol considered here, the quality of the correlations depends on two parameters: the ``covariance'' $c$ and the variance $v$ of Bob's states. To define these numbers in the case of a collective attack, we write the classical-quantum (cq) state shared by Alice and Bob in the PM version of the protocol as $\rho_{cq} = \frac{1}{4} \sum_{k=0}^3\Pi_k \otimes \mathcal{E}(|\alpha_k\rangle \langle \alpha_k|)$ where $\{\Pi_k\}_{k=0\ldots 3}$ are four orthogonal projectors and $\mathcal{E}$ denotes the quantum channel from Alice to Bob. Let us further define the quadrature operators on Bob's phase space as $\hat{q} = b + b^\dagger$, $\hat{p} = i(b^\dagger-b)$, with $b$ and $b^\dagger$ the annihilation and creation operators so that $[\hat{q}, \hat{p}] = 2i$. With these notations, we define
\begin{align}
\label{eqn:cv}
c&= \tr \left[\left( (\Pi_0-\Pi_2) \otimes \hat{q} + (\Pi_1-\Pi_3) \otimes \hat{p}\right) \rho_{cq}\right] \nonumber \\
v &= \frac{1}{2} \tr \left[\left( \mathbbm{1}_4 \otimes (\hat{q}^2+ \hat{p}^2)\right) \rho_{cq}\right]. 
\end{align}
As an example, we can compute these two parameters if the quantum channel between Alice and Bob is a bosonic phase-invariant Gaussian channel of transmittance $T$ and excess noise $\xi$, meaning that a coherent state $|\beta\rangle$ is mapped to a thermal state centered on $\sqrt{T} \beta$ with variance $1 + T \xi$. In this case, we obtain $c(T, \xi) = 2\sqrt{T} \alpha$ and $v(T, \xi) = 1 + 2T\alpha^2 + T\xi$.
In particular, under the assumption that the channel is Gaussian, one can recover the values of $T$ and $\xi$ from the parameters $c$ and $v$ observed in the protocol.

As already mentioned, the QPSK protocol presents a number of advantages against protocols with a Gaussian modulation of coherent states such as \cite{GG02, WLB04}. First, Alice simply needs to generate random bits and not Gaussian random variables that would then need to be discretized with sufficient precision. Second, the state preparation only requires a phase modulator, instead of both a phase and an amplitude modulators. Another strong argument in favor of this protocol is relative to the complexity of classical error correction. It is indeed well known that the reconciliation of Gaussian variables (as required for the protocols of \cite{GG02} and \cite{WLB04}) is quite costly and requires to decode classical error correcting codes of length $2L$ \cite{JKL11, JK12, JEK14}. In contrast, the binary nature of the raw key in the QPSK protocol allows Alice and Bob to aggregate the symbols in large blocks of size $m$, and to only decode classical codes of length $2L/m$, thus reducing the postprocessing complexity by a factor $m$ (which typically scales like $1/T$).

Of course, the QPSK protocol also has some limitations. In particular, for our security proof to provide a meaningful bound on the secret key rate, the mixture of four coherent states should approximate a thermal state, which limits the possible value of $\alpha$ to low numbers. A natural solution to this problem is to increase the size of the constellation and rely on more general QAM, as discussed in Section \ref{sec:larger}.

%%%%%%%%%%%%%%%%%%
\section{Challenges raised by a discrete modulation}
\label{sec:chal}

Establishing the security of CVQKD against general attacks turned out to be much more challenging that for BB84-like protocols. Currently, there exist two main approaches to do so. The first approach relies on an entropic uncertainty principle and has been successfully applied to the protocol of Ref.~\cite{CLV01} which requires Alice to prepare squeezed states \cite{FFB12}. For the moment, it is unclear whether a tighter version of entropic uncertainty principle could also work for protocols with coherent states (see Ref.~\cite{CBT17} for a review). The second approach follows a general strategy for establishing the security of a protocol against general attacks: one first appeals to a de Finetti-type theorem to reduce the problem to the case of collective attacks, and security against collective attacks is analyzed thanks to some version of the asymptotic equipartition property \cite{TCR09}, stating essentially that the asymptotic secret key rate is given by the so-called Devetak-Winter rate $K_{\mathrm{DW}}$ \cite{DW05}:
\begin{align}
\label{eqn:dw}
K_{\mathrm{DW}} = I (X;Y) - \sup \chi(Y;E),
\end{align}
where $I(X;Y)$ stands for the mutual information between Alice's variable $X$ and Bob's variable $Y$, $\chi(Y;E)$ for the Holevo information between $Y$ and Eve's quantum system $E$, with the supremum computed over all quantum channels $\mathcal{E}$ compatible with the correlations $c$ and $v$ observed during parameter estimation. 
Before providing more details about $K_{\mathrm{DW}}$, let us mention that de Finetti-type theorems exist for continuous-variable systems: Ref.~\cite{RC09} provides a (rather loose) version valid for permutation-invariant protocols (which is the case of essentially all CVQKD protocols), and Ref.~\cite{lev17} gives a tighter version but only for protocols displaying a stronger invariance in phase-space, such as the protocols of \cite{WLB04, POS15}.
Studying collective attacks, \textit{i.e.}, computing $K_{\mathrm{DW}}$, is rather straightforward for BB84-like protocols since it only involves an optimization over some finite-dimensional space. However, this is not the case for CV protocols, and bounding the quantity $\sup \chi(Y;E)$ is nontrivial since one must optimize over states in the full Fock space. 
In fact, there are two different issues here: $(i)$ how to obtain a robust estimate of $c$ and $v$ defined in Eq.~\eqref{eqn:cv}; and $(ii)$ how to compute the supremum of $\chi(Y;E)$ over all states compatible with $c$ and $v$. 

Let us first examine the first question. For the moment, the only protocols for which we are able to analyze parameter estimation (of a covariance matrix), with the proper error bounds, are those with the invariance in phase-space, using the ideas of Ref.~\cite{lev15}. The difficulty is that the parameters to be estimated are not bounded (contrary to the case of BB84 where the error rate is between 0 and 1), and computing a confidence region for them requires that the protocol is invariant under unitary transformations in phase-space, or requires some additional assumptions (for instance that the state is Gaussian, or that some moment of the variables are upper bounded by some explicit value). In the present paper, we do not address this question that we leave for future work.

In order to discuss the second question, we need to be more precise about the term $\chi(Y;E)$. This Holevo information is computed for a tripartite quantum state $\rho_{AYE}$, which is a quantum-classical-quantum state obtained when Bob measures system $B$ of another state $\rho_{ABE}$ with heterodyne detection. 
These states appear in the entanglement-based (EB) version of the QPSK protocol. In this version, Alice initially prepares $L$ copies of the bipartite pure state $|\Phi\rangle = \frac{1}{2} \sum_{k=0}^3 |\psi_k\rangle_{A} |\alpha_k\rangle_{A'}$, where $\{|\psi_k\rangle\}_{k=0\ldots 3}$ forms an orthonormal basis of the space spanned by the four coherent states (the precise definition of $|\psi_k\rangle$ does not matter at this stage), keeps register $A$ and sends register $A'$ to Bob through the quantum channel. Note that if Alice measures register $A$ in the basis $\{|\psi_k\rangle\}_{k=0\ldots 3}$, then she projects the state in $A'$ onto one of the four coherent states, with uniform probability. Hence, the EB and PM versions of the protocol are undistinguishable from the outside of Alice's labs, which implies that both protocols have the same security.
In the context of a collective attack, it makes sense to describe the quantum channel between Alice and Bob by a completely positive trace preserving (CPTP) map: $\mathcal{E}: A' \to B$, or equivalently by an isometry $\mathcal{U}_{A' \to BE}$. 
The tripartite state shared by Alice, Bob and Eve then reads: 
\begin{align*}
\rho_{ABE} = (\mathrm{id}_A \otimes \mathcal{U}_{A' \to BE}) (|\Phi\rangle\langle \Phi|),
\end{align*}
where $\mathrm{id}_{A}$ is the identity map on register $A$. Register $B$ is then measured with heterodyne detection, which is modeled by another CPTP map $\mathcal{M}_{B \to Y}$, corresponding to the resolution of the identity by coherent states: $\mathbbm{1} = \frac{1}{\pi} \int_{\mathbbm{C}} |\alpha\rangle \langle \alpha|\, \mathrm{d}\alpha$. This finally gives $\rho_{AYE} = (\mathrm{id}_A \otimes\mathcal{M}_{B \to Y} \otimes \mathrm{id}_E )(\rho_{ABE})$.
One can also apply the isometry $\mathcal{U}_{A' \to BE}$ to the cq state $\rho_{\mathrm{cq}}^0 = \frac{1}{4}\sum_{k=0}^3 \Pi_k \otimes |\alpha_k\rangle \langle \alpha_k|$ and recovers $\rho_{cq} = \tr_E (\mathcal{U}_{A' \to BE}(\rho_{\mathrm{cq}}^0 ))$.
We are now ready to define the term $\sup \chi(Y;E)$ appearing in the Devetak-Winter rate: this is the supremum of the Holevo information between $Y$ and $E$ computed for $\rho_{AYE}$, optimized over all isometries $\mathcal{U}_{A' \to BE}$ yielding parameters $c$ and $v$ when applied to $\rho_{\mathrm{cq}}^0$.
In other words, Alice and Bob observe correlations in the PM protocol (corresponding to the version that they indeed implement in practice) and must infer a bound on $\chi(Y;E)$ computed on the tripartite state that they would share with Eve if they had instead implemented the EB version of the protocol. This bound should hold for any quantum channel compatible with the parameters they observe.

The optimization appearing in the Devetak-Winter is thus highly nontrivial for CV protocols since the isometry $A' \to BE$ is an arbitrary isometry between infinite-dimensional Fock spaces. Quite remarkably, it is possible to compute the supremum of $\chi(Y;E)$ over states $\rho_{AYE}$ with a fixed covariance matrix for $\rho_{AB}$. This is known as the optimality of Gaussian states \cite{WGC06}. A second remarkable fact is that when the modulation of coherent states is Gaussian in the PM version, then one can directly compute the covariance matrix of $\rho_{AB}$ from the correlations observed in the PM version \cite{GCW03}, and we will provide a short proof of this fact in Section \ref{sec:larger}. By combining both properties, one can then compute the Devetak-Winter rate for protocols involving a Gaussian modulation of coherent states \cite{GC06} (see also \cite{NGA06} for an alternative proof).

In the case of CV protocols with a discrete modulation, the optimality of Gaussian states still works and provides a bound on $\chi(Y;E)$ for a given covariance matrix of the state $\rho_{AB}$ appearing in the EB version of the protocol. What is missing, however, is a direct way to compute this covariance matrix from the parameters $c$ and $v$ accessible in an experiment (in the PM protocol). Solutions to this problem are to restrict the possible quantum channels to linear bosonic channels, as done in Ref.~\cite{LG09}, or to add decoy states as in \cite{LG11}. Neither solution is satisfactory, since the first does not yield a general security proof, and the second basically renders moot all the advantage of the discrete modulation (since Alice must still implement a Gaussian modulation, and the error correction procedure remains quite heavy).
We will now present a much better solution to this problem.

%%%%%%%%%%%%%%%%%%

\section{A lower bound in the asymptotic limit}
\label{sec:bound}

As we already pointed out, we do not consider composability issues in this work and in particular, we restrict our attention to the asymptotic scenario, assuming that the parameters $c$ and $v$ of Eq.~\eqref{eqn:cv} are known. Our goal is then to compute the Devetak-Winter rate $K_{\mathrm{DW}}$ of Eq.~\eqref{eqn:dw}. As explained in the previous section, thanks to the optimality of Gaussian states \cite{WGC06}, our task is simply to perform an optimization over the possible covariance matrices of $\rho_{AB}$ compatible with the values of $c$ and $v$.

We first discuss the special case of the pure-loss (noiseless) channel, before moving to the general case of arbitrary channels and providing some numerical results.

%%%%%%%%%%%%%%%%%%

\subsection{The pure-loss channel}
\label{subsec:loss}

Dealing with a pure-loss channel is much easier than the general case  because the pure-loss channel is essentially the only channel yielding parameters of the form $c = 2\sqrt{T}\alpha$ and $v = 1+ 2 T\alpha^2$ for some $T \in [0,1]$. Here, $\alpha$ is the amplitude of the coherent states prepared by Alice. 
From such parameters, one immediately infers that a coherent state $|\alpha_k\rangle$ is mapped to another coherent $|\sqrt{T}\alpha_k\rangle$. 
Without loss of generality, the isometry $\mathcal{U}$ is of the form $\mathcal{U} | \alpha_k\rangle_{A'}= |\sqrt{T} \alpha_k\rangle_B |\mu_k\rangle_E$ for some states $\{|\mu_k\rangle\}_{k=0\ldots 3}$. The output states have to be product states otherwise the output in register $B$ would not be pure and the channel would add some noise. Recall that the Gram matrix of a vector of states $(|v_1\rangle, \ldots, |v_n\rangle)$ is the $n \times n$ matrix $G$ with entries $G_{k, \ell} = \langle v_k, v_\ell\rangle$. We can see that the Gram matrices  of $\{|\sqrt{1-T} \alpha_k\rangle\}$ and $\{|\mu_k\rangle\}$ coincide, since $\langle \alpha_k |\alpha_\ell\rangle =\langle t \alpha_k |t\alpha_\ell\rangle\langle \mu_k |\mu_\ell\rangle=\langle t \alpha_k |t\alpha_\ell\rangle\langle \sqrt{1-T} \alpha_k |\sqrt{1-T} \alpha_\ell\rangle$, with $t = \sqrt{T}$. The first equality follows from the fact that $\mathcal{U}$ is an isometry and the second is obtained by applying a beam-splitter transformation of transmittance $T$.
Using the polar decomposition, if two Gram matrices of the form $M_1 M_1^\dagger$ and $M_2 M_2^\dagger$ coincide, then there exists some isometry $V$ such that $M_1 = M_2 V$. In particular, it means that there is a local isometry mapping $|\mu_k\rangle$ to $|r\alpha_k\rangle$, with $r = \sqrt{1-T}$. 
This proves that the channel can also be modeled as 
\begin{align}
\mathcal{U}' |\alpha_k\rangle_{A'}  = |\sqrt{T} \alpha_k\rangle_B |\sqrt{1-T} \alpha_k\rangle_E,
\end{align}
and therefore that the channel behaves like the pure-loss channel restricted to our set of states. 
In particular, since we know the value of of $c$ and therefore of $T$, it is easy to compute the covariance matrix of $\rho_{AB}$ in the EB version of the protocol.

%%%%%%%%%%%%%%%%%%

\subsection{General lower bound via semidefinite programming}
\label{subsec:sdp}

We now turn to the general case of dealing with noisy channels. 
Let us recast our problem by considering the EB version of the protocol. Alice prepares the initial state
\[ |\Phi\rangle := (\mathbbm{1} \otimes \sqrt{\rho_{\mathrm{PM}}}) |\mathrm{EPR}\rangle,\]
where $\rho_{\mathrm{PM}} := \frac{1}{4} \sum_{k=0}^3 |\alpha_k\rangle \langle \alpha_k|$ is the mixture of the 4 coherent states prepared in the PM protocol and $|\mathrm{EPR} \rangle := \sum_{n=0}^\infty |n,n\rangle$ is the maximally entangled (unnormalized) state between two modes. This state is a purification of $\rho_{\mathrm{PM}}$ and this specific choice is made because it maximises the correlation between its two modes.
More explicitly, we obtain:
\begin{align*}
|\Phi\rangle = \frac{1}{2} \sum_{k=0}^3 |\psi_k\rangle |\alpha_k\rangle,
\end{align*}
with $|\psi_k\rangle = \frac{1}{2} \sum_{m=0}^3 e^{-i km \frac{\pi}{2}} |\phi_m\rangle$ and 
\begin{align*}
|\phi_m\rangle &= \frac{1}{\sqrt{\nu_m}} \sum_{n=0}^\infty  \frac{\alpha^{4n+m}}{\sqrt{(4n+m)!}}|4n+m\rangle
\end{align*}
where $\nu_{0,2} = \frac{1}{2}(\cosh(\alpha^2) \pm \cos(\alpha^2))$, $\nu_{1,3} = \frac{1}{2}(\sinh(\alpha^2) \pm \sin(\alpha^2))$ and $|4n+m\rangle$ denotes the Fock state with $4n+m$ photons.

The quantum channel between Alice and Bob can be described \textit{via} its Kraus operators $\{E_i\}$ which satisfy $\sum_i E_i^\dagger E_i = \mathbbm{1}_{A'}$. The quantum state $\rho_{AB}= ( \mathrm{id}_A \otimes \mathcal{E})(|\Phi\rangle \langle \Phi|)$ is therefore 
\begin{align}
\rho_{AB} = \frac{1}{4} \sum_{k, \ell=0}^3 |\psi_k\rangle \langle \psi_\ell| \otimes \sigma_{k, \ell},
\end{align}
where we defined $\sigma_{k, \ell} = \sum_i E_i |\alpha_k\rangle \langle \alpha_\ell | E_i^\dagger$.

Our goal is to bound the covariance matrix of $\rho_{AB}$ for any possible quantum channel $\mathcal{E}$ yielding some fixed values for $c$ and $v$. By symmetry of the protocol, we are in fact only interested in 3 parameters, corresponding to the variance of $\rho_A$, the variance of $\rho_B$ and the covariance. This means that without loss of generality, we can assume that the covariance matrix  takes the following form $\left[ \begin{smallmatrix} V_A \mathbbm{1}_2 & Z \sigma_Z \\ Z \sigma_Z & V_B \mathbbm{1}_2\end{smallmatrix} \right]$, where $\sigma_Z = \left[\begin{smallmatrix} 1 & 0 \\ 0 & -1 \end{smallmatrix} \right]$, and $V_A = 1 + 2\alpha^2$ only depends on $|\Phi\rangle$, $V_B = v$. In particular, there is a single unknown, $Z$, that we need to bound. Since $\chi(Y;E)$ is a decreasing function of $Z$ when the other parameters are fixed, we only need to get a lower bound on $Z$ as a function of $c$ and $v$. 
The parameter $Z$ is defined as the expectation of $\frac{1}{2}(\hat{q}_A \hat{q}_B- \hat{p}_A \hat{p}_B)$ for the state $\rho_{AB}$, which corresponds to
\begin{align*}
Z  = \tr \left[( ab + a^\dagger b^\dagger) \rho_{AB} \right],
\end{align*}
where $a$ and $a^\dagger$ are the annihilation and creation operators on the Fock space of register $A$. 

Let us define $\Pi = \sum_{k=0}^3 |\psi_k\rangle \langle \psi_k|$ to be the orthogonal projector onto the space spanned by the four coherent states and $C = \Pi a \Pi \otimes b + \Pi a^\dagger \Pi \otimes b^\dagger$. With these notations, we have $Z = \tr(C X)$where $X$ is the (unknown) state $\rho_{AB}$. 
This matrix $X$, which is positive semidefinite with trace 1, must satisfy some linear constraints, namely $\tr(B_0 X)= v$ and $\tr( B_1 X) = c$ for
\begin{align*}
B_0 &= \Pi \otimes (1 + 2 b^\dagger b)\\
B_1 &= \left( (|\psi_0\rangle \langle \psi_0|-|\psi_2\rangle\langle \psi_2|) \otimes \hat{q} \right.\\
&\left. \quad + (|\psi_1\rangle \langle \psi_1| - |\psi_3\rangle\langle \psi_3|) \otimes \hat{p}\right).
\end{align*}
The final constraint is $\tr_B X = \tr_B |\Phi\rangle \langle \Phi |$, which is $\sum_{k,\ell=0}^3 \langle \alpha_\ell | \alpha_k\rangle \, |\psi_k\rangle \langle \psi_\ell|$.
In other words, we are interested in the following problem:
\begin{align}\label{eqn:sdp}
\min & \, \tr( C X) \\
\text{such that} &\left\{
\begin{array}{l}
\tr(B_0 X) =v \nonumber \\
 \tr( B_1 X) = c \nonumber \\
 \tr( B_{k,\ell} X ) = \frac{1}{4} \langle \alpha_\ell | \alpha_k\rangle \nonumber \\
  X  \succeq 0, \nonumber
\end{array}
\right.
\end{align}
where the last constraint means that $X$ is positive semidefinite and where we have defined $B_{k, \ell} = |\psi_\ell\rangle\langle \psi_k|$. 
This is a semidefinite program which can be solved numerically.
Denoting by $Z^*$ the optimum of this program, we are able to compute an explicit lower bound on $\sup \chi(Y;E)$ by taking the value of the Holevo information for a Gaussian state $\rho_{AB}^*$ with covariance matrix $\Gamma^* = \left[ \begin{smallmatrix}(1+ 2\alpha^2) \mathbbm{1}_2 & Z^* \sigma_Z \\ Z^* \sigma_Z & v \mathbbm{1}_2\end{smallmatrix} \right]$. 
This quantity is then computed with standard techniques \cite{LBG07} and is given by
\begin{align*}
\chi(Y;E)_{\rho_{AB}^*} = g\left(\frac{\nu_1-1}{2}\right)+ g\left(\frac{\nu_2-1}{2}\right)- g\left(\frac{\nu_3-1}{2}\right),
\end{align*}
where $g(x) := (x+1) \log_2 (x+1) - x \log_2(x)$, $\nu_1$ and $\nu_2$ are the symplectic eigenvalues of $\Gamma^*$ and $\nu_3 = 1+2\alpha^2 - \frac{Z^{*2}}{1+v}$. It satisfies $\chi(Y;E)_{\rho_{AB}^*} \geq \sup_{\mathcal{U}_{A'\to BE}} \chi(Y;E)$ where the optimization is over isometries compatible with parameters $c$ and $v$.
 We present numerical results in the next subsection.

One might wonder whether all the solutions of this program correspond to valid quantum states for some quantum channel $\mathcal{E}$. This is the case since the only constraint that must be satisfied by any channel is that $\tr_B X = \tr_B |\Phi\rangle \langle \Phi |$. In other words, because the initial state is pure, and because all purifications of $\rho_A$ are equivalent up to an isometry on the purifying system $BE$, there always exists an isometry from $A'$ to $BE$ mapping $|\Phi\rangle$ to any valid solution $X$ of the SDP.

%%%%%%%%%%%%%%%%%%

\subsection{Numerical results}
\label{subsec:rate}

In this section, we compute the key rate for Gaussian channels characterized by a transmittance $T$ and excess noise $\xi$. It is important to note that the proof presented above does not make any assumption about the quantum channel $\mathcal{E}$ between Alice and Bob, since the mutual information between their data, as well as the values of $c$ and $v$ can be estimated during the protocol. In order to display numerical results without sampled data, we will use the expressions of $I(X;Y)$, $c$ and $v$ as functions of $T$ and $\xi$,  as given for Gaussian channels which provide a realistic model for quantum channels that typically occur in experiments. The values computed from the SDP will thus give lower bounds for the key rates, easy to compare to the ones assuming a Gaussian or linear channel \cite{LG09}. 
To take into account the imperfect error correction procedure between Alice and Bob, as in realistic implementations, we plot a modified version of the Devetak-Winter rate given by $\beta I(X;Y) - \sup \chi(Y;E)$, with a reconciliation efficiency parameter $\beta \leq 1$.
The mutual information $I(X;Y)$ should be computed for a \textit{binary-input} additive white Gaussian noise (AWGN) channel \footnote{The capacity of the binary AWGN channel is 
\unexpanded{$C_{\text{bi-AWGN}}(s) = -  {\int} {\phi_s}(x)  {\log}_2({\phi_s}(x)) \mathrm{d}x + \frac{1}{2} \log_2\left(\frac{s}{2\pi e}\right) $}
where {$\phi_s(x) = \sqrt{\frac{s}{8\pi}}\left(e^{-s(x+1)^2/2} + e^{-s(x-1)^2/2}\right)$} and $s$ is the signal-to-noise ratio, and the capacity of the AWGN channel is {$C_{\text{AWGN}}(s) = \frac{1}{2} \log_2 (1+s)$}. For $s \leq 0.5$, which is the case here since $s = 2T\alpha^2/(2+T\xi)$, the two capacities are essentially equal.}, but in the relevant regime of parameters for us, it is very well approximated by the capacity of an AWGN channel and given by 
\begin{align*}
I(X;Y) \approx \log_2 \left(1 + \frac{2T \alpha^2}{2+T \xi}\right).
\end{align*}
For each channel, we compute the parameters $c(T, \xi)$ and $v(T, \xi)$ that Alice and Bob would obtain during parameter estimation (in the asymptotic limit), and solve the SDP of Eq.~\eqref{eqn:sdp} to upper bound $\sup \chi(Y;E)$ by some $\chi(Y;E)_{\rho_{AB}^*}$. Since this SDP involves infinite-dimensional matrices, it is necessary to truncate this space in order to get numerical results. 
It is natural to truncate the Fock space of Bob by the space spanned by the first $N$ Fock states: $|0\rangle, |1\rangle, \ldots, |N-1\rangle$, thus obtaining a full Hilbert space of dimension $4N$ (since Alice's local space can be taken to be the 4-dimensional space spanned by $\{|\alpha_k\rangle\}_{k=0 \ldots 3}$). 
In practice, we observe that the results do not depend on the specific value of $N$ provided that it is larger than 10. Note that the fact that we need to truncate the Fock space is not necessarily an important issue for security proofs: this is because composable security proofs of CVQKD usually require to project the state onto a low-dimensional subspace of the Fock space anyway, via some energy test \cite{RC09}.  We use the solver SCS \cite{ocpb:16,scs} and set the precision below $10^{-5}$. 

\begin{figure}[!h]
\centering
  \includegraphics[width=1\linewidth]{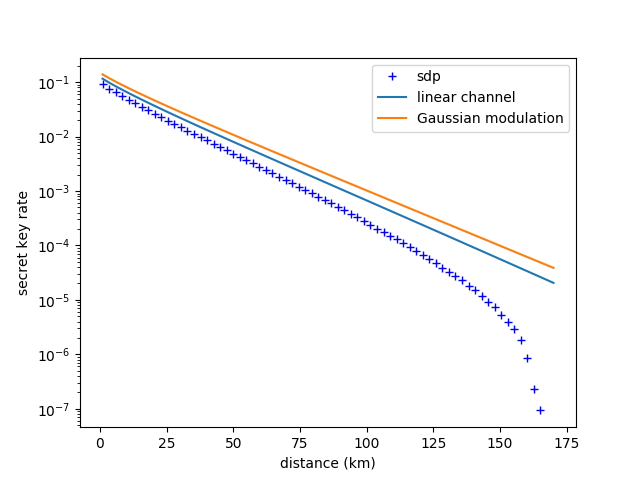}  
\caption{Secret key rate versus distance, for a Gaussian channel with transmittance $T=10^{-0.02 d}$ and excess noise $\xi = 0.002$. Here $d$ is the distance between Alice and Bob in km. The value of $\alpha$ is $0.35$. The reconciliation efficiency $\beta$ is set to $0.95$. The top curve corresponds to the performance of the protocol \cite{WLB04} with a Gaussian modulation, the lower curve to the performance of the four-state protocol while assuming a linear channel (as in \cite{LG09}) and the crosses correspond to the lower bound given by our SDP.  }
  \label{fig:2}
\end{figure}

\begin{figure}[!h]
\centering
  \includegraphics[width=1\linewidth]{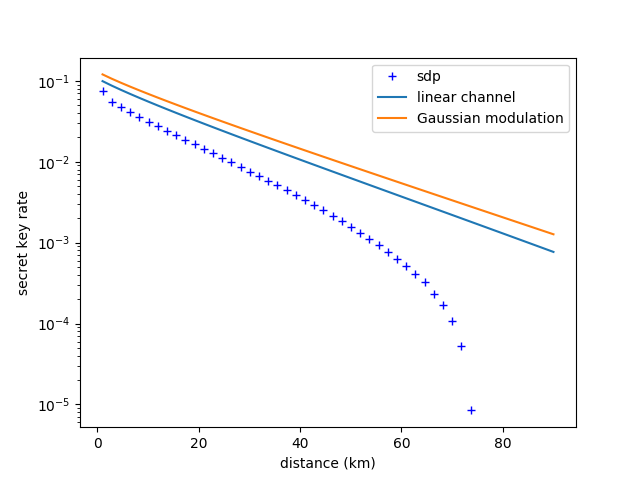}
\caption{Secret key rate versus distance, for a Gaussian channel with transmittance $T=10^{-0.02 d}$ and excess noise $\xi = 0.005$. Other parameters are the same as in Fig.~\ref{fig:2}.}  
  \label{fig:5}
\end{figure}

\begin{figure}[!h]
\centering
  \includegraphics[width=1\linewidth]{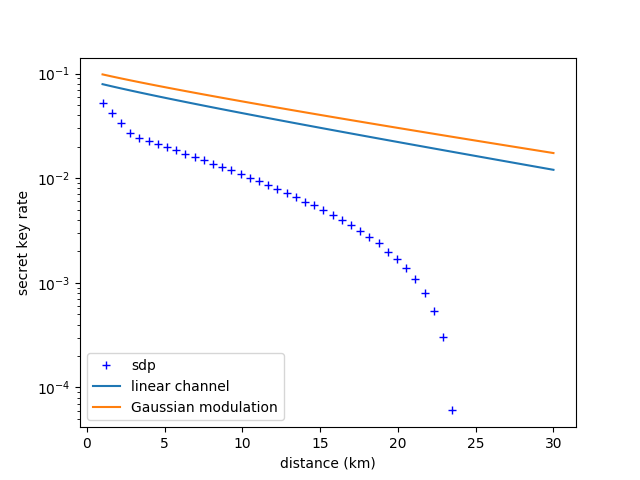}  
\caption{Secret key rate versus distance, for a Gaussian channel with transmittance $T=10^{-0.02 d}$ and excess noise $\xi = 0.01$. Other parameters are the same as in Fig.~\ref{fig:2}.}  
  \label{fig:10}
\end{figure}

We plot our lower bound on the Devetak-Winter rate
\begin{align*}
\beta \log_2 \left(1 + \frac{2T \alpha^2}{2+T \xi}\right) - \chi(Y;E)_{\rho_{AB}^*} \leq K_{\mathrm{DW}}
\end{align*}
for three different values of excess noise: $\xi=0.002$ in Fig.~\ref{fig:2}, $\xi=0.005$ in Fig.~\ref{fig:5} and $\xi=0.01$ in Fig.~\ref{fig:10}. We remark that distances much larger than 100 km are possible provided that the excess noise is sufficiently small, and that such values have already been obtained in experimental demonstrations \cite{JKL13, HIM17}. 
Note that in realistic implementations, the detectors are inevitably noisy and display a limited efficiency. In a scenario where these imperfections are possibly controlled by the eavesdropper, the secret key rate would be much lower than the ones displayed on Fig.~\ref{fig:2}, \ref{fig:5} and \ref{fig:10}. It is, however, legitimate to consider a more optimistic scenario where the imperfections of the detectors are not assumed to be controlled by the eavesdropper \cite{SBC08}. In this case, the secret key rate can be computed following the method of \cite{LBG07}. Because the effect of imperfections in the trusted-detector-noise scenario is typically quite mild \cite{UF16}, we chose to ignore it here and assumed ideal detectors for Bob.

\begin{figure}[!h]
\centering
  \includegraphics[width=1\linewidth]{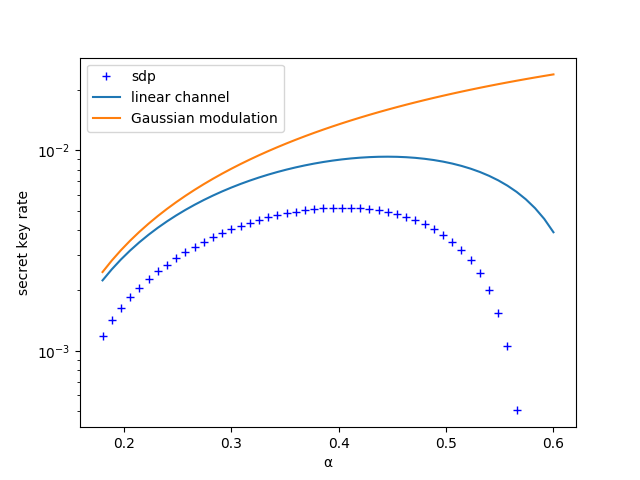}  
\caption{Secret key rate versus $\alpha$, for a distance of $50$ km and excess noise of $\xi=0.002$. Other parameters are the same as in Fig.~\ref{fig:2}.}  
  \label{fig:alpha}
\end{figure}

As we noted earlier, the main limitation of the QPSK protocol probably concerns the small value of $\alpha$. Indeed, our approach relies on the closeness between a thermal state (corresponding to the Gaussian modulation, and for which we know the exact secret key rate) and a mixture of four coherent states. These two mixtures are only approximately undistinguishable in the regime where $\alpha \ll 1$, and indeed the performance of the QPSK protocol degrades rapidly for $\alpha \geq 0.5$, corresponding to about $\alpha^2 \approx 0.25$ photon per pulse. This behaviour is illustrated in Fig.~\ref{fig:alpha}.
To overcome this limitation, it is possible to exploit more complicated quadrature amplitude modulations (QAM) that will better approximate thermal states with a large variance, as discussed below. This is notably explored in \cite{JKD12, KGW19} in the context of quantum key distribution and in \cite{LRS17,GJR17,JEM18} for communication over bosonic Gaussian channels. 

%%%%%%%%%%%%%%%%%%

\section{Larger constellations}
\label{sec:larger}

While we chose to illustrate our technique with the QPSK modulation in this paper, our SDP approach generalizes in a straightforward way to more complex modulation schemes. 
For these schemes, we start with a target Gaussian modulation, described by some thermal state 
$$\rho(\gamma) = (1- \gamma^2) \sum_{k=0}^\infty \gamma^{2k} |k\rangle \langle k|$$
 of parameter $\gamma >0$, and a good modulation scheme will aim at approximating this state by a mixture of a finite number of coherent states.

Consider for instance a modulation where $n$ coherent states $\{|\alpha_k\rangle\}_{k =1 \ldots n}$ are prepared with probability $\{p_k\}_{k =1 \ldots n}$. A possible example would be to take $n$ coherent states on a circle (Phase-Shift Keying) of the form $|\alpha e^{i k \frac{2\pi}{m}}\rangle$, as considered for instance in \cite{SL10, PLW18}, or more general QAM as in Ref.~\cite{LZG18}. 
The average state prepared by Alice in the PM version is simply $\rho_n = \sum_{k=1}^n p_k |\alpha_k\rangle \langle \alpha_k|$.
In the EB version of the protocol, Alice would prepare the initial bipartite pure state $|\Phi_n \rangle = (\mathbbm{1} \otimes \sqrt{\rho_n}) \sum_{i=0}^\infty |i\rangle |i\rangle$
where $|i\rangle$ is a Fock state with $i$ photons. This specific choice of purification is made to maximize the value of the parameter $Z$ in the covariance matrix, and therefore to maximize the resulting lower bound on the secret key rate.
In particular, the objective function of our SDP will be $\tr \left( (ab +a^\dagger b^\dagger) \rho_{AB}\right)$ with $\rho_{AB} =( \mathrm{id} \otimes \mathcal{E} ) (|\Phi_n\rangle\langle\Phi_n|)$.

We now need to write the constraints of our SDP. The first constraint is simply that the partial trace $\tr_{B} (\rho_{AB})$ should coincide with the partial trace of the initial state, $\tr_B (|\Phi_n\rangle \langle \Phi_n|) = \rho_n$. This yields the constraint $\tr_B (\rho_{AB}) = \rho_n$. The second constraint corresponds to the variance of Bob's reduced state, and this is given as before by $\tr( \mathbbm{1}\otimes (\mathbbm{1} + 2 b^\dagger b) X) = v$.
The third constraint requires slightly more work since one needs to relate the correlations $c$ observed in the PM protocol to a measurement applied to $\rho_{AB}$. 

For a general QAM, the best way to define $c$ is similar to what is done in the protocols with a Gaussian modulation: it should be the average of the dot product between the $L$-dimensional complex vector $(\alpha_{k_1}, \ldots, \alpha_{k_L})$ of states sent by Alice and the $L$-dimensional complex vector $(\beta_1, \ldots, \beta_L)$ of measurement results of Bob. Here $\beta_{\ell}$ is the outcome of the heterodyne detection of $\mathcal{E}(|\alpha_{k_\ell}\rangle\langle \alpha_{k_\ell}| )$, which is the state received by Bob for the $\ell^{\mathrm{th}}$ use of the channel. This dot product can be alternatively written as the expectation of $\bar{\alpha_k} \beta_k$, where the conjugation is a consequence of working with complex variables. 
Let us denote by $M_\infty^1$ the observable corresponding to heterodyne detection:
$$M_\infty^1 = \frac{1}{\pi} \int_{\mathbbm{C}} \beta |\beta \rangle \langle \beta| \mathrm{d} \beta.
$$
Our definition of $c$ is therefore:
$$
c:= \sum_{k=1}^n p_k \bar{\alpha}_k \mathrm{tr} (M_\infty^1 \mathcal{E}(|\alpha_k\rangle \langle \alpha_k|)).
$$
We now need to express $c$ as the expectation of an observable applied to the state $\rho_{AB}$ in the EB protocol.
First, we observe that by construction, there exists an $n$-outcome measurement $\{F_k\}_{k =1 \ldots n}$ on system $A$ such that outcome $k$ prepares the coherent state $|\alpha_k\rangle$ on the second mode. To see this, let us introduce the purification $|\Phi'\rangle_{CB} = \sum_{k=1}^n \sqrt{p_k} |\phi_k\rangle_C |\alpha_k\rangle_B$ of $\rho_n$, where $\{|\phi_k\rangle\}_{k=1\ldots n}$ is an orthonormal family. Both $|\Phi_n\rangle_{AB}$ and $|\Phi'\rangle_{CB}$ are purifications of $\rho_n$ so there exists an isometry $V: C\to B$ such that $(V \otimes \mathbbm{1})|\Phi'\rangle_{CB} = |\Phi_n\rangle_{AB}$ and one can choose $F_k = V |\phi_k\rangle \langle \phi_k| V^\dagger$.
This measurement satisfies $\sum_{k=1}^n F_k= \mathbbm{1}$ and $\langle \Phi_n| F_k \otimes \mathbbm{1} |\Phi_n\rangle = p_k$. 
Let us define the following complex-valued observable $M_n = \sum_{k=1}^n \alpha_k  F_k$. It correctly yields $\alpha_k$ when the state sent by Alice through the quantum channel is $|\alpha_k\rangle$.
We can finally use the fact that $\tr_A (M_n^\dagger \rho_{AB}) = \sum_{k=1}^n p_k \bar{\alpha}_k \mathcal{E}(|\alpha_k\rangle\langle \alpha_k|)$ to express $c$ as:
$$ c=\tr( (M_n^\dagger \otimes M_\infty^1) \rho_{AB}).$$

With these notations in place, we are now ready to define the SDP that computes the term $\tr \left( (ab + a^\dagger b^\dagger) \rho_{AB}\right)$ of the covariance matrix of $\rho_{AB}$ in the EB version of the protocol, namely 
\begin{align}
\label{eqn:sdpn}
\min & \quad \tr((ab+ a^\dagger b^\dagger) X)\\
\text{such that} &\left\{
\begin{array}{l}
 \tr_B X = \rho_n\nonumber\\
\tr( \mathbbm{1}\otimes (\mathbbm{1} + 2 b^\dagger b) X) = v \nonumber\\
 \tr( (M_n^\dagger \otimes M_\infty^1) X) = c\nonumber\\
 X \succeq 0.\nonumber
\end{array}
\right.
\end{align}
 The final constraint simply expresses that $X$ (corresponding to our unknown state $\rho_{AB}$) is a valid density matrix, hence a positive semidefinite operator. Just as before, the solution $Z^*$ of this program yields a covariance matrix $\Gamma^* = \left[ \begin{smallmatrix} V_A \mathbbm{1}_2 & Z \sigma_Z^* \\ Z^* \sigma_Z & V_B \mathbbm{1}_2\end{smallmatrix} \right]$, where $\sigma_Z = \left[\begin{smallmatrix} 1 & 0 \\ 0 & -1 \end{smallmatrix} \right]$ and $V_A$ is now the variance of $\rho_n$, which can be used to compute the upper bound $ \chi(Y;E)_{\rho_{AB}^*}$ on the Holevo information between Bob and Eve. 

Such an SDP can be solved efficiently, but its size appears to grow quite rapidly with the number $n$ of states in the constellation. This is because the state $\rho_{AB}$ is represented by an $nN \times nN$ matrix, with $n$ the dimension of Alice's space (spanned by $n$ coherent states) and an $N$-dimensional truncation of Bob's Fock space.
For large constellations, a better idea might be to truncate Alice's Hilbert space to the first $N$ Fock states, which would yield a matrix of size $N^2 \times N^2$.

It is instructive to consider what happens in the limit $n \to \infty$ where the constellation becomes exactly Gaussian.
In that case, the observable $M_n$ tends to the (rescaled and conjugated) heterodyne detection $(M_\infty^\gamma)^\dagger = \frac{1}{\pi} \int_{\mathbbm{C}} \gamma \bar{\beta} |\beta \rangle \langle \beta| \mathrm{d} \beta$ as the constellation approaches the thermal state $\rho(\gamma)$:
\begin{align*}
M_n \xrightarrow[\rho_n \to \rho(\gamma)]{} (M_\infty^\gamma)^\dagger.
\end{align*}
This is because the purification $(\mathbbm{1} \otimes \sqrt{\rho(\gamma)}) \sum_{i=0}^\infty |i\rangle |i\rangle = \sqrt{1- \gamma^2} \sum_{k=0}^\infty \gamma^k |k\rangle|k\rangle$ of a thermal state $\rho(\gamma)$ is a two-mode squeezed vacuum state and performing a heterodyne detection (corresponding to $M_\infty^1$) on the first mode prepares a coherent state $|\gamma \bar{\alpha}\rangle$ for the second mode upon measurement result $\alpha$.
In that case, the third constraint becomes $\tr(( (M_\infty^\gamma)^\dagger \otimes M_\infty^1)  X)= c$. We also know that a heterodyne detection is nothing but two noisy homodyne detections, which gives
\begin{align*}
\tr(( (M_\infty^\gamma)^\dagger \otimes M_\infty^1)  X) &= \gamma \, \tr(( (M_\infty^1)^\dagger \otimes M_\infty^1)  X)\\
&= \gamma \, \tr\left(\frac{1}{2} (\hat{q}_A \otimes \hat{q}_B - \hat{p}_A \otimes \hat{p}_B)  X\right)\\
&= \gamma \, \tr( (a b + a^\dagger b^\dagger)  X).
\end{align*}
In other words, the objective function of the SDP is simply a scalar multiple of the third constraint. As a consequence, the solution is unique and given by $\gamma^{-1} c$, which is indeed the correct value of the covariance for a CVQKD protocol with Gaussian modulation \cite{GCW03}.

Since the limit of the SDP for large constellations ($n \to \infty$) recovers the value of the secret key rate for protocols with a Gaussian modulation, it is tempting to exploit continuity arguments to show that the secret key rate of CVQKD protocols with large constellations is close to that of Gaussian protocols. To make this quantitative, one must study the stability of the SDP of Eq.~\eqref{eqn:sdpn} against small perturbations in the constraints, namely when $\rho_n$ approximates $\rho(\gamma)$ and $M_n$ approximates $M_\infty^\gamma$ in the first and third constraints, respectively. Such questions have been studied in the literature on complex optimization, for instance in Ref.~\cite{BS00}, but are beyond the scope of the present work. 

%%%%%%%%%%%%%%%%%%
\section{Discussion and perspectives}
\label{sec:disc}

In this work, we give a general technique to derive a lower bound on the secret key rate of CVQKD with a discrete modulation, and apply it to the case of the QPSK modulation. We do not expect this bound to be tight and believe that it could likely be improved, but this would require fundamentally new proof techniques. The bound is loose because it crucially relies on Gaussian optimality, meaning that $\chi(Y;E)$ is computed for the Gaussian state with the same covariance matrix as the one returned by the SDP. That state, however, is non Gaussian, and $\chi(Y;E)$ is therefore overestimated. This is clear for instance in the QPSK protocol because $\rho_{A}$ is a mixture of four coherent states and therefore non Gaussian. The issue is that the SDP is not looking for a state that would yield the maximum value of $\chi(Y;E)$ but rather for a state with a very specific covariance matrix. At the same time, this restriction disappears when the size of the constellation increases since the SDP bound converges to the optimal secret key rate in the limit of a Gaussian modulation.

A remaining open question in the field of CVQKD is whether one can provide a composable security proof against general attacks for protocols with a discrete modulation. 
We do not get such a composable security proof here because we do not analyze the parameter estimation procedure. While parameter estimation is rather straightforward for BB84-like protocols, the situation is more complicated for continuous variables because we need to obtain a confidence region for parameters such as the variance of Bob's state which are unbounded. Because of that, standard statistical tools to get tail bounds on distributions of random variables such as the Chernoff bound or variants do not apply anymore. A solution is to exploit some specific symmetry of the protocol in phase-space as in Ref.~\cite{lev15}, but discrete modulations break this symmetry and a new approach is therefore needed. 
At the same time, the fact that Bob's detection is rotationally invariant gives hope that a rigorous analysis of the parameter estimation procedure should be possible. Combining such an analysis with our results would then yield a composable security proof valid against collective attacks, and the exponential de Finetti theorem of Renner and Cirac would then imply a composable security proof valid against general attacks \cite{RC09}, albeit with pessimistic bounds in the finite-size regime. 
This points to two important questions open to future work: analyze the parameter estimation procedure of protocols with a discrete modulation, and improve on the exponential de Finetti theorem of Ref.~\cite{RC09}.

%%%%%%%%%%%%%%%%%%

\section{Conclusion}

In this work, we focus on the CVQKD protocol with a QPSK modulation and establish a lower bound on its secret key rate in the asymptotic limit. 
This bound is obtained by solving a semidefinite program that computes the covariance matrix of the state shared by Alice and Bob in the entanglement-based version of the protocol. While our bounds are likely not tight, they already show that secret key can be distributed over more than 100 km for realistic values of the excess noise.
We also show how the same technique can be applied to analyze the security of more complicated QAM. This is a major step towards the full security of CVQKD with a discrete modulation. If the parameter estimation procedure of such protocols could be analysed rigorously, our result would imply a composable security proof valid against general attacks. We leave this question for future work.

%%%%%%%%%%%%%%%%%%

\newpage

\acknowledgements{We thank Yann Balland for discussions at the early stage of this project. We acknowledge funding from European Union's Horizon's Horizon 2020 Research and Innovation Programme under Grant Agreements No.~675662 (QCALL) and  No.~ 820466 (CiViQ), and from the French National Research Agency (ANR) project quBIC.}

%\nocite{apsrev41Control}
%\bibliographystyle{ieeetr}
%\bibliography{CV}

\end{document}